\documentclass[conference]{IEEEtran}
\IEEEoverridecommandlockouts
\usepackage{geometry}
\usepackage{cite}
\usepackage{amsmath,amssymb,amsfonts}
\usepackage{graphicx}
\usepackage{textcomp}
\usepackage[noend]{algpseudocode}
\usepackage{algorithm}
\usepackage{url}

\usepackage{xcolor}
\def\BibTeX{{\rm B\kern-.05em{\sc i\kern-.025em b}\kern-.08em
    T\kern-.1667em\lower.7ex\hbox{E}\kern-.125emX}}
\usepackage{listings}
\usepackage{xcolor}

\begin{document}

\title{ Real-Time Control and Monitoring of Photovoltaic Arrays Using RTDS and BeagleBoard Technology
\\
\thanks{Md~Fazley~Rafy is with the LCSEE, WVU, doing his Ph.D. in Computer Science. Corresponding author: Md~Fazley~Rafy (E-mail: mr00065@mix.wvu.edu)}
}

\author{Md~Fazley~Rafy}

\maketitle
\begin{abstract}
Increasing integration of alternate electricity generation due to declining fossil fuels is becoming an essential option for future generations. As photovoltaic technology brings forth enormous benefits to the alternate solution for future power grid systems, this paper presents a comprehensive real-time system designed to model, control, and monitor a PV array subjected to dynamic loads using the Real-Time Digital Simulator (RTDS) environment. Integration with the Generic Transducer Network (GTNET)- Socket(SKT) module allows for the simulation of various environmental conditions, such as changes in insolation and temperature, and their direct impact on the PV array's performance. Utilizing BeagleBoard technology, the system demonstrates the capability to modify these conditions through real-time data input, subsequently observing the effects on current and voltage output curves. The real-time simulation results are visualized as a SCADA system within the Real-time Simulation for Automated Controller Design (RSCAD) runtime environment, providing insights into the effective management of solar power systems.

\end{abstract}


\section{Introduction}

The increase in environmental pollution and scarcity of fossil fuels is raising concerns with the exponential growth of technology and development in the 21st Century. A net-zero initiative of governments worldwide has led to the consideration of alternate energy solutions, such as renewable generation. Renewable generation includes power and energy generated from energy conversion from heat, mechanical power, water flow, wind, biomass gas emission, etc. The resulting technologies that facilitate such conversion to generate essential energy-based electricity include PV solar arrays, wind turbines, battery storage systems, bio-mass gas emitters, and many more. Among these, the prevalent one is solar power conversion through PV to generate solar-based electricity. United States Department of Energy reported to have generated 5,259,595 MWh electricity by renewable energy of which 31.5\% was from Solar PV alone \cite{doe2022}. The inclusion of PV-based sustainable energy resources is widely used due to its ubiquity, abundance from sunlight, and sustainability. The source of this technology is the sole impact of solar energy from sunlight, which makes it free of cost and desirable. As the PV arrays are made of semiconductor materials, the integrated devices are static, complete, and liberated from moving parts. These characteristics make a solar panel easy to maintain, operate, and scale. The operation of solar panels generally relies on solar irradiance, the temperature of the solar cells, operating voltage, surrounding temperature, dust, humidity, wind intensity, etc. Among all these factors, insolation and temperature are the most critical ones that influence the output power and voltage of the panels which proportionately influence the generation of electricity \cite{ASIM20125834}. This paper considers these two factors to control the output voltage and current of the PV arrays that provide electricity to the loads, which can be small residential or large commercial facilities. However, with the increasing integration of PV array-based distributed energy resources (DERs), the evolving landscape of modern power systems is becoming more complex and multimodal \cite{benigni2020real}. Energy management of such systems requires effective control of the solar operations to maintain grid stability and meet consumer demands. This complexity necessitates a comprehensive testing of different system components. Real-time simulation with a hardware-in-the-loop (HIL) can validate the performance of these intricate mechanisms without concern of losing essential power or damaging expensive components in PV panels. The need for an efficient HIL-based real-time simulation is the core focus of this report which dictates the control of multiple factors that facilitate the PV to function constructively. There are different simulation tools to assess the performance of solar PV, and comparison between PVGIS, PV Watts, and PV Syst dictates how real-time PV data monitoring can increase efficiency \cite{thotakura2020operational}. 
\begin{figure*}[t!]
    \centering
    \includegraphics[width=1.9\columnwidth]{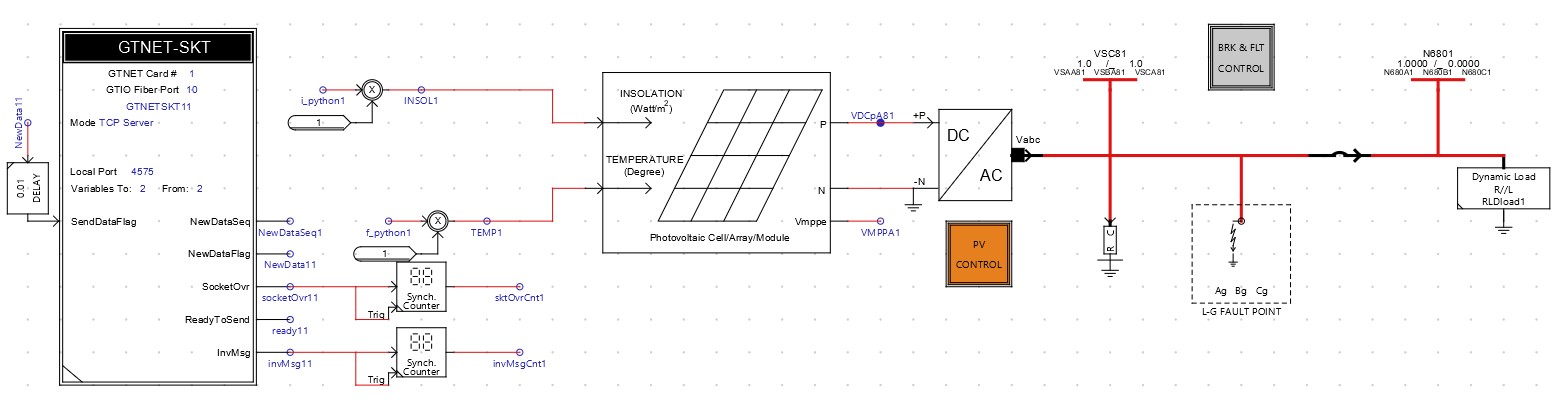}
    \caption{RSCAD system design with PV, controller, GTNET-SKT, and Load}
    \label{fig:design}
\end{figure*}
Digital simulators like RTDS and Opal-RT are also used to validate the performance of the sliding mode controller and the Perturb \& Observe (P\&O) controller for a solar panel \cite{kolluru2015real}. RTDS perform power flow and other computation at the microsecond level to simulate a real-time environment for PV arrays that requires dynamic behavior representation under various conditions. This real-time simulation is a useful resource for education and training on power electronics for solar PV system integration. The Opal-RT on the other hand is another competitive simulator due to its ability to offer a flexible platform for experimentation on all aspects of a PV-integrated system \cite{jiang2017real}. Both of these provide a practical and efficient way to develop and test multiple scenarios in real time, which is crucial for both research and educational purposes in the rapidly evolving field of solar PV systems. To include the HIL in the Rea; Time System (RTS), black BeagleBoards are used as a controller for PV parameters. Among different types of BeagleBoards, the black version is the most common one as it has both wired and wireless capabilities. In Internet-of-Things (IoT) based technology, these boards are used considering their powerful processing capabilities, suitability, and usability to implement ModBus protocol, competence to manage OPC USA server-client, and integration capability with diverse IoT applications \cite{ventuneac2021implementation}. Regardless of the type, BeagleBoards are used for academic support for their cost-effectiveness, versatility in software and hardware integration, and other appropriateness in teaching data analysis and acquisition. These boards, like PocketBeagle, offer pragmatic solutions, and experimental capabilities for learning embedded systems and programming, which is an ideal device in technical fields \cite{strom2020embedded}. Compared to Arduino Boards or Raspberry Pi, black BeagleBoards are better for control systems for PV as they can manage more extensive data with complex algorithms, making them suitable for sophisticated monitoring devices. However, their higher cost and potential limitations in processing power for very large-scale data need to be considered for larger implementation \cite{barbosa2019beaglebone}. To objectify the necessity of efficient PV control for beneficial power system management of DERs, this paper scratches the surface of the control mechanism to improve PV performance. Section \ref{method_sec} presents the details of the design, operation, and data flow of the model. Software and Hardware setup for the experiment in RTS is explained in Section \ref{des_sec}. SCADA performance and illustration of the control through discussion are shown in Section \ref{res_sec}. Finally, the overall contribution and summary of the paper are in Section \ref{conclusion_sec}.


\section{Methodology}\label{method_sec}

The core operation for this RTS is guided by the RSCAD design as shown in Figure \ref{fig:design}. GTNET-SKT card and Beagle boards, as explained in Section \ref{des_sec} further articulate the RTS process with hardware inclusion to control the parameters of the PV array. With the varying measurement for insolation and temperature with varying environmental conditions, a PV controller, as shown in Figure \ref{fig:design} uses this information to adjust the operation of the PV array. The adjustment provides optimization of output power under those varying different environmental conditions. BeagleBoards are acting as two factors of the environment that are changing with time and affecting the performance of the PV. The complete steps of RTS operation with PV and Beagle boards are demonstrated in Algorithm 1 for better exposition.

\begin{algorithm}\label{algo}
\caption{Control and Monitoring of PV Array System}
\begin{algorithmic}

\State Initialize RSCAD Environment
\State Set up PV Array Model with dynamic load in RTDS
\State Integrate GTNET-SKT Module for inputting isolation and temperature

\State Initialize BeagleBoard 1
\State Run Linux
\State Connect to Router and Network
\State Run Python Script for Insolation Control
\While{True}
    \State Get Insolation Value
    \State Send Insolation Value to RTDS via Socket Communication
\EndWhile

\State Initialize BeagleBoard 2
\State Run Linux
\State Connect to Router and Network
\State Run Python Script for Temperature Control
\While{True}
    \State Get Temperature Value
    \State Send Temperature Value to RTDS via Socket Communication
\EndWhile

\State Monitor and Control Loop
\While{True}{
    \State Receive Insolation Value from BeagleBoard 1
    \State Receive Temperature Value from BeagleBoard 2
    \State Update PV Array Model in RTDS with new Insolation and Temperature
    \State Generate and Update Current and Voltage Output Curves in Real-Time
    \State Display Output Curves as SCADA in RSCAD Runtime Environment}
\EndWhile

\end{algorithmic}
\end{algorithm}
\subsection{RTDS Modeling}\label{}
RTDS is a powerful simulator to realize the real-time behavior and dynamics in power systems through design in a proprietary software environment, referred to as RSACD. Control logic, scripting and automation, integration, and operation on external devices by interacting with RTDS are the most common applications of RSCAD. For our HIL experiment, RSCAD provides a base software model to simulate the PV control and operations in real-time. As shown in Figure \ref{fig:design}, the components of the model include a GTNET-SKT module, a PV array, a DC-AC converter, a dynamic R/L load, and physical wiring. GTNET-SKT is used to connect remote external devices through socket communication. The module accepts certain particular values of a defined number of variables and sends them to the PV array as an input, which is in our case real-time values for temperature and insolation. The solar-powered array acts as a generator that produces DC power and the converter transforms the DC to the required AC power that is ideal for the connected dynamic load. The load consists of Resistance and Inductance to exhibit dynamic or time-varying behavior of power consumption. The converter contains a Voltage Source Converter (VSC) that controls the DC link voltage to provide active power to the load and maintains synchronization. A delta-wye transformer is also modeled inside the converter box to step up the VSC AC-side voltage to load-level voltage and reduce harmonic current injection to the grid. Capacitor filters and interface reactors inside the converter box also are fitted at the point of common coupling (PCC) between the VSC and the dynamic load to reduce the harmonics injection of the converter. Moreover, a PV control, a Breaker, and a Fault control are included in the design to connect the PV array to the load to maintain power quality and safety in the PCC.

\begin{figure}[htbp]
    \centering
    \includegraphics[width=\columnwidth]{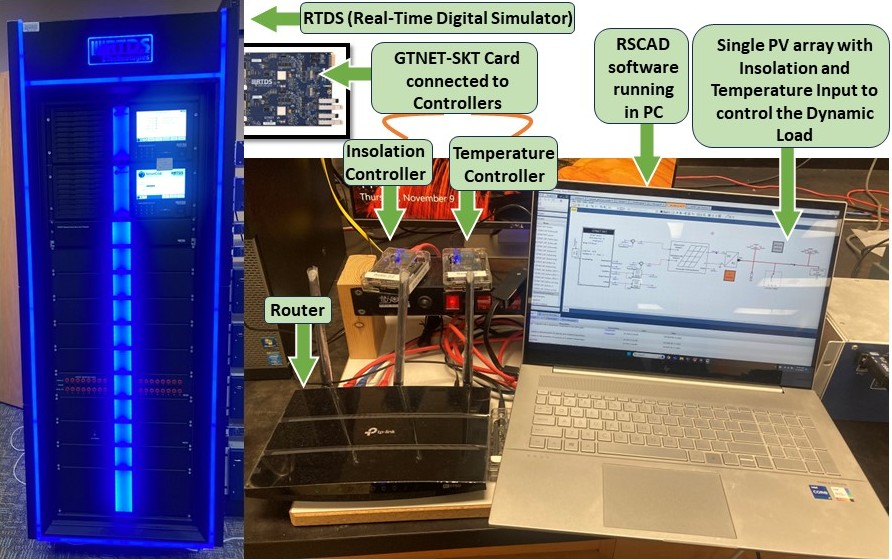}
    \caption{HIL setup with BeagleBoard controllers and RTDS}
    \label{fig:HIL}
\end{figure}

\subsection{Real-Time Control Setup}\label{control_sec}
The RTS control was designed in this model with the integration of the GTNET-SKT module as a software tool, however, it is a hardware card that physically connects the external devices to the RTDS as shown in Figure \ref{fig:HIL}. The GTNET-SKT card accepts wired fiber or Ethernet connections from the external network components and enables remote communication capabilities in RTDS \cite{chen2014implementing}. Inside the RSCAD, the card is configured to work as a TCP server with transmitting and receiving capabilities. There are two types of ports that need to be specified: a fiber port number that physically represents the port of the GTNET-Card and another socket port, which is 4575, to connect client devices. The GTNET-SKT module is also configured to hold two floating number variables (i\_python1 and f\_python1), representing insolation (INSOL1) and temperature values (TEMP1) of the solar panel, to store the received values from the clients as illustrated in Figure \ref{fig:design}. Two counters are also included in the design to log the frequency of communication between RTDS and BeagleBoards.

\subsection{BeagleBoard Implementation}\label{bgl_sec}

Two Beagleboard black are used in the HIL experiment as a controller for solar power generators in RTDS. The boards are equipped with a specific processor which is powered by an ARM Cortex-A8 microprocessor. Providing a balance between power consumption and processing power, BeagleBoard black includes Synchronous Dynamic Random-Access Memory (SDRAM), which is a single 256Mb$X16$ DDR3L with 4Gb (512MB) memory device and Flash memory with an external memory card. Ubuntu is installed as an operating system on the board and it has a USB connector to access the console to run essential programs. In addition to the USB, the board has 10/100 Ethernet capability for wired communication and can be powered through 5V DC input from the USB. This device with PowerVR SGX530 graphics accelerator and 1 GHz clock speed based on ARM processor makes it a reliable and powerful controller device for PV input control \cite{he2014use}. A custom python script, as provided in Appendix 1 as pseudocode, is run through the console to communicate and send insolation and temperature values at fixed time intervals to the RSCAD PV model. The Python script enables socket communication to connect with the GTNET-SKT server and control the insolation and temperature to simulate an RTS environment.

\section{System Design}\label{des_sec}

With the help of RTDS, the real-time system is modeled as the simulator is able to perform computation with synchronized clock time. The larger number of input-output channels and dynamic processing power enable experiments with hardware-in-the-loop simulations. Facilitating that function, beagleboards were possible to integrate with the software design in RTDS through a special transducer, GTNET-SKT, that enables a versatile interface between different external communication devices and system models in RSCAD. As demonstrated in Figure \ref{fig:HIL}, RTDS is controlled through RSCAD software, and PV design and controller are modeled in the RSCAD. Inside the model, the GTNET-SKT module refers to the actual hardware device in RTDS, which is also known as the GTNET card, and connects the router to create a local network. In that local network, beagle boards are also connected and can establish communication through the GTNET card with the help of a router. The details of the device's configurations and performance monitoring of the overall system are provided in the subsequent sections.

\subsection{Hardware Architecture}
Major components of the HIL setup include RTDS, a router, two BeagleBoard Black, and a laptop to show RSCAD design and SCADA in the runtime environment. The router connects the beagleboards to the same local network as the RTDS that has a GTNET-SKT card as the network component with proper MAC and IP addresses. The Beagleboards send values to the GNTET card which is received by the GTNET-SKT module to feed the PV array modeled in RSCAD in the laptop to control power generation and load consumption.

\begin{figure}[htbp!]
    \centering
    \includegraphics[width=\columnwidth]{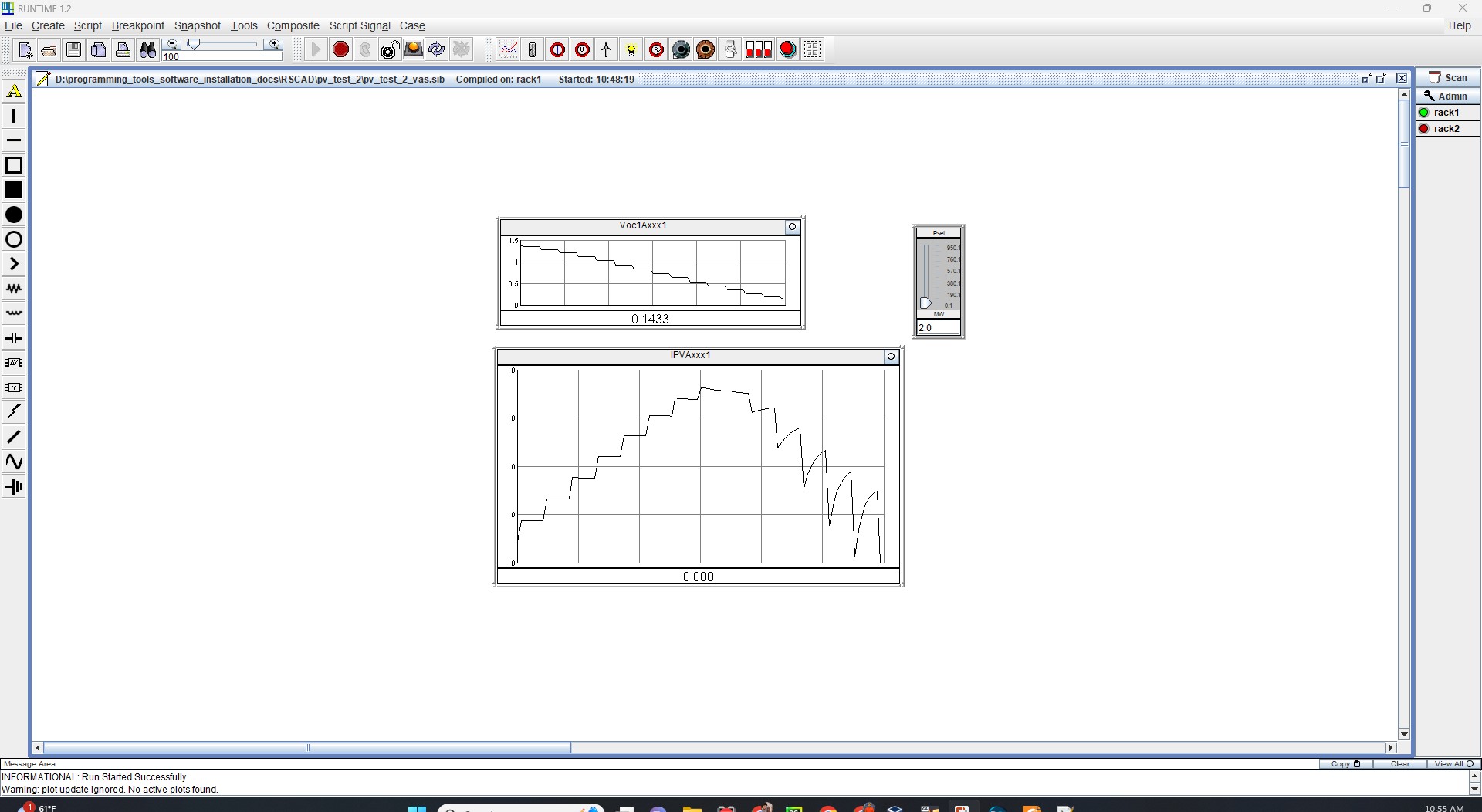}
    \caption{SCADA demonstration with Changing I-V curve with alternating Insolation and Temperature}
    \label{fig:SCADA}
\end{figure}
\subsection{Software Architecture}
RSCAD Runtime, a powerful real-time simulation and control platform, can be effectively utilized as a SCADA (Supervisory Control and Data Acquisition) system for monitoring and controlling various aspects of a power system. In our experiment, we have illustrated current and voltage graphs as shown in Figure \ref{fig:SCADA} as the changes in response to varying insolation and temperature conditions. Additionally, a slider control is shown in the Figure that introduces the dynamic behavior of the load with varying values to relate the realistic consumption of power between 5W to 30W.

\section{Results and Discussion}\label{res_sec}
The SCADA results shown in Figure \ref{fig:SCADA}, demonstrate the effect the environmental conditions like insolation and temperature on solar panel performance. The use of RTDS as a digital simulator allows for the modeling, control, and monitoring of PV arrays with connected dynamic loads to realize the RTS. The GTNET-SKT further assist in the control of these factors through the controllers to adjust the different rate of insolation and degree of temperature throughout the day. The SACADA shows how current and voltages change over time. The current changes proportionately with the increasing insolation and constant temperature. While the increase is less during temperature changes with constant insolation. With varying loads, and increasing insolation high flow of power is observed from the solar panel. However, improved efficiency is observed with lower temperatures and higher insolation that can help meet varying load demands more effectively.

The HIL-based PV controller in RTS demonstrates a significant advancement in the field of renewable energy management and monitoring, particularly in photovoltaic (PV) systems, by leveraging Real-Time Digital Simulation (RTDS) integrated with BeagleBoard technology. The developed system, as outlined through Algorithm 1, offers several key commercial and technical benefits that are crucial in today's energy landscape:

\subsection{Enhanced Efficiency and Reliability} The real-time simulation and monitoring of PV arrays ensure optimal performance under varying environmental conditions. This system's ability to dynamically adjust to changes in insolation and temperature results in more efficient and reliable energy production, which is vital for large-scale solar power plants and distributed solar energy systems.

\subsection{Predictive Maintenance and Reduced Downtime} With continuous monitoring and immediate feedback on PV array performance, potential issues can be identified and addressed proactively. This predictive maintenance approach reduces the downtime and extends the lifespan of solar panels, thereby maximizing return on investment for solar energy projects.

\subsection{Scalability and Flexibility for Grid Integration} The modular nature of the system, facilitated by the use of BeagleBoard microcontrollers, makes it highly scalable and adaptable to various sizes of PV installations. This flexibility is essential for integrating solar power into the existing energy grid, especially as the grid evolves to accommodate more renewable energy sources.

\subsection{Cost-Effectiveness in Energy Management} The utilization of cost-effective BeagleBoard microcontrollers and the efficiency gains from the system can significantly reduce operational costs. This cost-effectiveness is particularly beneficial for commercial solar energy providers, making renewable energy more competitive with traditional energy sources.

\subsection{Educational and Research Applications} The system provides a valuable tool for educational and research purposes, offering a practical platform for studying and experimenting with PV systems. This can foster innovation and contribute to the development of more advanced solar energy technologies.

\subsection{Support for Sustainable Energy Goals} By improving the efficiency and reliability of PV systems, this project aligns with global initiatives towards sustainable energy and reduced carbon footprint. It supports the transition to clean energy, which is crucial for environmental preservation and combating climate change.

\section{Conclusion}\label{conclusion_sec}

Future smart grid technology with the increasing integration of PV arrays demands efficient analysis and operation of PV controllers to balance the demand from load consumption and response from the solar power generator.
This paper demonstrates the impact of varying environmental factors, such as insolation and temperature to realize the need for proper PV control mechanisms for efficient power consumption and generation. RTDS and Beagle boards were used to develop a HIL testing environment to validate the performance of solar power generators under the influence of different external environmental factors. The real-time control and monitoring system for PV arrays presented in this paper not only advances the technical capabilities of solar energy management but also offers tangible benefits for commercial applications, research, and education. This system stands as a testament to the potential of integrating modern control technologies into renewable energy systems, paving the way for a more sustainable and efficient energy future.

\bibliographystyle{unsrt}
\bibliography{term_ref.bib}

\begin{thebibliography}{10}

\bibitem{doe2022}
Federal Energy~Management Program.
\newblock Federal agency use of renewable electric energy.
\newblock \url{https://www.energy.gov/femp/federal-agency-use-renewable-electric-energy#:~:text=Federal%20Renewable%20Electricity%20Requirement&text=%C2%A7%2015852%20)%2C%20each%20fiscal%20year,as%20the%20renewable%20electricity%20requirement.}
\newblock Accessed: 2023-11-9.

\bibitem{ASIM20125834}
Nilofar Asim, Kamaruzzaman Sopian, Shideh Ahmadi, Kasra Saeedfar, M.A. Alghoul, Omidreza Saadatian, and Saleem~H. Zaidi.
\newblock A review on the role of materials science in solar cells.
\newblock {\em Renewable and Sustainable Energy Reviews}, 16(8):5834--5847, 2012.

\bibitem{benigni2020real}
Andrea Benigni, Thomas Strasser, Giovanni De~Carne, Marco Liserre, Marco Cupelli, and Antonello Monti.
\newblock Real-time simulation-based testing of modern energy systems: A review and discussion.
\newblock {\em IEEE industrial electronics magazine}, 14(2):28--39, 2020.

\bibitem{thotakura2020operational}
Sandhya Thotakura, Sri~Chandan Kondamudi, J~Francis Xavier, Ma~Quanjin, Guduru~Ramakrishna Reddy, Pavan Gangwar, and Sri~Lakshmi Davuluri.
\newblock Operational performance of megawatt-scale grid integrated rooftop solar pv system in tropical wet and dry climates of india.
\newblock {\em Case Studies in Thermal Engineering}, 18:100602, 2020.

\bibitem{kolluru2015real}
Venkata~Ratnam Kolluru, Kamalakanta Mahapatra, and Bidyadhar Subudhi.
\newblock Real-time digital simulation and analysis of sliding mode and p\&o mppt algorithms for a pv system.
\newblock {\em International Journal of Emerging Electric Power Systems}, 16(4):313--322, 2015.

\bibitem{jiang2017real}
Zhenning Jiang, Georgios Konstantinou, Zhaoyu Zhong, and Pablo Acuna.
\newblock Real-time digital simulation based laboratory test-bench development for research and education on solar pv systems.
\newblock In {\em 2017 Australasian Universities Power Engineering Conference (AUPEC)}, pages 1--6. IEEE, 2017.

\bibitem{ventuneac2021implementation}
Cornel Ventuneac and Vasile~Gheorghita Gaitan.
\newblock Implementation of an iiot access gateway for the modbuse--modbus extension using beaglebone black.
\newblock {\em EIRP Proceedings}, 16(1), 2021.

\bibitem{strom2020embedded}
Stephen~A Strom and Marius Strom.
\newblock Embedded measurement and control applications utilizing python on the pocket beaglebone.
\newblock In {\em 2020 ASEE Virtual Annual Conference Content Access}, 2020.

\bibitem{barbosa2019beaglebone}
Yaniel~Sousa Barbosa, Alexandre~Zambujo Brito, Luis~Brito Palma, Paulo~Sousa Gil, and Rui~Azevedo Antunes.
\newblock Beaglebone black for x8-vb quadcopter attitude control.
\newblock In {\em IECON 2019-45th Annual Conference of the IEEE Industrial Electronics Society}, volume~1, pages 311--317. IEEE, 2019.

\bibitem{chen2014implementing}
Bo~Chen, Karen~L Butler-Purry, Ana Goulart, and Deepa Kundur.
\newblock Implementing a real-time cyber-physical system test bed in rtds and opnet.
\newblock In {\em 2014 North American Power Symposium (NAPS)}, pages 1--6. IEEE, 2014.

\bibitem{he2014use}
Nannan He, Han-Way Huang, and Brian~David Woltman.
\newblock The use of beaglebone black board in engineering design and development.
\newblock In {\em 2014 ASEE North Midwest Section Conference}, volume~1. University of Iowa, 2014.

\end{thebibliography}

\end{document}